\documentclass[aps,prl,twocolumn,showpacs,groupedaddress]{revtex4}
\usepackage{graphicx}
\begin{document}

\title{
       Structural and microscopic relaxation processes
       in liquid Hydrogen Fluoride
      }

\author{
       R.~Angelini$^{1}$, P.~Giura$^{1}$,
       G.~Monaco$^{1}$, G.~Ruocco$^{3}$, F.~Sette$^{1}$,
       and R.~Verbeni$^{1}$.
}
\affiliation{
         $^{1}$ European Synchrotron Radiation Facility.
         B.P. 220 F-38043 Grenoble, Cedex France.\\
         $^{3}$ Universit\'a di Roma "La Sapienza" and Istituto Nazionale
         di Fisica della Materia, I-00185, Roma, Italy.\\
}
\date{\today}
\begin{abstract}

The high frequency collective dynamics of liquid hydrogen fluoride
(HF) is studied by inelastic x-ray scattering on the coexistence
curve at $T$=239 K. The comparison with existing molecular
dynamics simulations shows the existence of two active relaxation
processes with characteristic time-scales in the sub-picosecond
range. The observed scenario is very similar to that found in
liquid water. This suggests that hydrogen bonded liquids behave
similarly to other very different systems as simple and glass
forming liquids, thus indicating that these two relaxation
processes are universal features of the liquid state.
\end{abstract}
\maketitle

A marked characteristic of the liquid state is that the dynamics
of the density fluctuations is controlled by relaxation processes.
For example, in liquids, the structural re-arrangement of the
particles takes place with a characteristic time-scale,
$\tau_\alpha$, dictated by the local inter-particles interaction
and the actual thermodynamic state. The presence of these
re-arrangements affects the dynamics of the density fluctuations
as it allows energy exchanges between different density
fluctuation modes. In this context, one calls {\it relaxation
process} the mechanism governing these energy flows, and, in this
specific example, we have described the well known {\it
structural- or $\alpha$-relaxation process}\cite{harrison}.
Other relaxation processes may be active,
each one characterized by a specific underlying physical
mechanism. One of the open problems in the physics of the liquid
state is to understand on a general ground the common features of
these relaxation processes. In simple monatomic liquids both
kinetic and mode-coupling theories predict the existence of two
distinct relaxation processes \cite{balucani} and, this prediction
has been demonstrated both by numerical simulations
\cite{PRA916671974, PRA2628591982} and by experiments
\cite{JPC1280092000}. In these theories, one of the two processes
is the $\alpha$-relaxation, and the other is a faster process
({\it microscopic} or {\it instantaneous} process) which is
thought to be associated with the interactions between an atom and
the ''cage'' of its nearest neighbors. Other relaxation processes,
beyond the $\alpha$ and the instantaneous processes, associated
with the internal molecular degrees of freedom may be observed in
molecular liquids. In glass-forming systems, one
also finds the same relaxation processes pattern. In this case, by
driving the system to the glassy state, where the structural
arrest freezes the $\alpha$-process and $\tau_\alpha$ becomes
extremely large, is still possible to clearly observe the
microscopic process \cite{PRE615872000}.

The outlined scenario seems to point that the $\alpha-$ and $\mu-$
relaxation processes are universal features of the liquid state.
In this context, however, no attempt has been made so far to
include the important class of hydrogen bonded liquids. This is
the aim of the present work.

The operational method to identify relaxation processes is the
study of the dynamic structure factor, $S(Q,\omega)$, and of its
inelastic features due to collective excitations - at momentum
transfer $Q$ these are observed at energy $\hbar\Omega(Q)$. The
dispersion relation of $\Omega(Q)$ allows to define an "apparent"
sound velocity, $c(Q)=\Omega(Q)/Q$, which is a constant in the low
Q-limit, and decreases with $Q$ approaching the inverse of the
inter-particle distance $d$. Whenever a relaxation process with
characteristic time $\tau$ is active, $c(Q)$ has a further
$Q$-dependence which shows up as a transition from a low frequency
value, $c_o$, to a high frequency one, $c_\infty$. This transition
takes place when the condition $\Omega(Q)\approx 1/\tau$ is
fulfilled. The amplitude of the jump between $c_\infty$ and $c_o$
defines the strength, $\Delta$, of the coupling between the
density fluctuations and the degrees of freedom involved in the
relaxation process. Multiple relaxation processes will give rise
to multiple "jumps" in $c(Q)$. Specifically, in the case of
$\alpha$- and $\mu$- relaxation processes, one expects that $c(Q)$
goes from $c_o$ to $c_{\infty\alpha}$ because of the
$\alpha$-process, and then from $c_{o\mu}= c_{\infty\alpha}$ to
$c_{\infty}$ because of the $\mu$-process \cite{nota1}. In real
systems, this picture is often an over-simplification. Indeed:
${\it i)}$ The sound velocities $c_o$, $c_{\infty\alpha}$ and
$c_{\infty}$, as well as the associated quantities $\tau$ and
$\Delta$ are $Q$-dependent, especially when $Qd\approx 1$. {\it
ii)} In certain cases the condition $\omega\tau_\mu=1$ cannot be
reached because $1/\tau_{\mu}(Q)$ is always larger than the
maximum value of $\Omega(Q)$.

In the framework of hydrogen bonded liquids, water has been
extensively studied and we report in Fig.~1 a summary of our
current understanding of the collective dynamics in terms of
apparent sound velocity. Experimental and numerical simulation
data points for ambient conditions fall between the $c_o(Q)$ and
$c_\infty(Q)$ curves determined by computer simulations. We
observe a positive dispersion of $c(Q)$ from $c_o(Q)$ ($\approx$
1500 m/s at $Q=0$) to a value of $\approx$ 3200 m/s in the $Q$
region $Q$=2$\div$4 nm$^{-1}$. Extensive temperature and density
dependent studies have shown that this positive dispersion is
associated to the $\alpha$-process \cite{harrison}. It is then
tempting to associate the large difference between
$c_{\infty\alpha}$ and $c_{\infty}(Q)$, e.~g. between
$\approx$3200 and $\approx$5000 m/s at $Q$=6 nm$^{-1}$, to the
$\mu$-process. This observation, evident in the low $Q$ region, is
less obvious at large $Q$ where the difference between
$c_{\infty\alpha}$ and $c_{\infty}(Q)$ is much smaller.

To clarify whether a $\mu-$process can be clearly identified in a
hydrogen bonded liquid we studied hydrogen fluoride. We report the
first experimental data of the $S(Q,\omega)$ of HF, and compare
them to existing simulation data \cite{PRL8120801998} . The
picture emerging from our analysis shows that in HF both the
$\alpha-$ and $\mu-$ process are active, and they can be clearly
identified because their relative strength is substantially weaker
than in liquid water. This study confirms, therefore, that these
two relaxation processes are active also in hydrogen bonded
liquids.

The $S(Q,\omega)$ of HF has been studied at T=239 K by Inelastic
X-ray Scattering (IXS) as a function of wave vector $Q$ in the
range 2 $\div$ 31 nm$^{-1}$. The HF sample ($\approx$ 1 $cm^3$)
was confined in a stainless steel cell equipped with two sapphire
windows (total thickness $500\  \mu m$) to allow the passage of
the incident and scattered x-rays. The experiment has been carried
out at the very high energy resolution IXS beam-line ID16 at the
European Synchrotron Radiation Facility. The instrument consists
of a back-scattering monochromator and five independent analyzer
systems, held one next to each other with a constant angular
offset on a 6.5 m long analyzer arm. The used Si(11 11 11)
configuration \cite{NIMPRB1173391996}, gives an instrumental
energy resolution of 1.6 meV full width half maximum (FWHM) and an
offset of 3 $nm^{-1}$ between two neighbor analyzers. The momentum
transfer is selected by rotating the analyzer arm. The spectra at
constant Q and as a function of energy were measured with a Q
resolution of $0.4\ \ nm^{-1}$ FWHM. The energy scans  were
performed varying the monochromator temperature with respect to
that of the analyzer crystals. Further details on the beam-line
are reported elsewhere \cite{NIMPRB1111811996}. Each scan took
about 180 min and each spectrum  at fixed Q was obtained by
summing up to 6 scans.

In Fig.~\ref{fig2}(a) we report selected spectra at the indicated
momentum transfer and the corresponding measured resolution
functions aligned and scaled to the central peak. The contribution
of the empty cell was found to be negligible. The spectra consist
of a band centered at zero energy transfer which becomes broader
with increasing $Q$ and with a characteristic asymmetry due to the
detailed balance. A preliminary determination of $c(Q)$ was
obtained fitting the spectra to the convolution of the resolution
function with a model function for $S(Q,\omega)$ composed by a
Lorentzian for the central peak ($S_o(Q,\omega)$)and a Damped
Harmonic Oscillator (DHO) \cite{ILLTR92FA008S1992} for the
inelastic signal ($S_i(Q,\omega)$). The deconvoluted inelastic
part of the current spectra, $\omega^2/Q^2S_i(Q,\omega)$, -whose
maxima correspond to the parameter $\Omega(Q)$- is reported in
Fig.2b. The dispersion curve $\Omega(Q)$ vs. $Q$ is shown in
Fig.~3 for the low $Q$ region data. For $Q$ between 4 and 7
nm$^{-1}$, it shows a linear dependence with a slope corresponding
to a sound speed of $1080$ m/s. This value is substantially higher
than the adiabatic sound speed $c_o=580$ m/s as obtained from
Brillouin light scattering \cite{dapubblicare}. Moreover, in the 2
to 4 nm$^{-1}$, the IXS data are compatible with a transition of
$c(Q)$ from its low frequency value $c_o$ to the higher value.
This increase of $c(Q)$ is then interpreted as due to the
$\alpha-$relaxation, and it comes out to be quite similar to that
of water, where the ratio $c_{\infty\alpha}/c_o$ was also close to
two. A more formal procedure to describe the effect of a
relaxation process in the $S(Q,\omega)$ is based on the
viscoelastic model. In this approach the $S(Q,\omega)$ is
expressed as:

\begin{equation}
S(Q,\omega)=I(Q){{\omega_0(Q)}^2 M^{\prime}(Q,\omega)\over
[\omega^2-\omega_0(Q)^2-\omega M^{\prime \prime}(Q,\omega)]^2
+[\omega M^{\prime}(Q,\omega)]^2}
\end{equation}

\noindent
where $\omega_0(Q)^2= (K_BT / m S(Q))Q^2$ is the
normalized second frequency moment of $S(Q,\omega)$, $K_B$ is the
Boltzmann constant, $m$ is the mass of the molecule and
$M^{\prime}(Q,\omega)$, $M^{\prime \prime}(Q,\omega)$ are
respectively the real and the imaginary part of the Laplace
transform of the memory function $M(Q,t)$. In the two relaxation
processes scenario, we model $M(Q,t)$ by the sum of two
exponential decay contributions:

\begin{equation}
M(Q,t)= \Delta_\alpha^2(Q)e^{-{t / \tau_\alpha (Q)}}
+ \Delta_\mu^2(Q)e^{-{t/ \tau_\mu(Q)}}
\end{equation}

\noindent where
$\Delta_\alpha^2(Q)=[c_{\infty\alpha}(Q)^2-c_0(Q)^2]Q^2$, and
$\Delta_\mu^2(Q)=[c_{\infty}(Q)^2-c_{\infty\alpha}(Q)^2]Q^2$ are
the strengths of the two processes. As, similarly to water, one
expects that the $\mu$-process is very fast with respect to the
investigated timescale \cite{PRE6055051999}, the second term is
approximated by a $\delta$-function:

\begin{equation}
M(Q,t)= \Delta_\alpha^2(Q)e^{-{t/ \tau_\alpha(Q)}} + \Gamma_\mu(Q)\delta(t)
\end{equation}

\noindent with $\Gamma_\mu(Q)= \Delta_\mu^2 \tau_\mu(Q)$. The
experimental data were fitted to the convolution of the
experimental resolution function with the dynamic structure
factor model given by Eq.~1. The relevant independent parameters
are $c_{\infty \alpha}(Q)$, $c_o(Q)$, $\tau_\alpha(Q)$ and
$\Gamma_\mu(Q)$. In Fig.~4 we show the values obtained for
$c_{\infty}(Q)$ and $c_0(Q)$, together with those for $c(Q)$ as
deduced from the DHO model. We observe that the positive
dispersion found for $c(Q)$ by the DHO analysis takes place
between the values of $c_o(Q)$ and $c_{\infty\alpha}(Q)$, derived
from the viscoelastic analysis. Therefore, this finding confirms
the hypothesis that the transition of $c(Q)$ is governed by the
$\alpha$- process. In particular, for $Q$ larger than 4 nm$^{-1}$,
the coincidence of $c(Q)$ and $c_{\infty\alpha}(Q)$ tells us that
the $\alpha$- process is fully accomplished for $Q$ larger than 4
nm$^{-1}$, and, in the explored $Q$ region, there is no evidence
for a further dispersion of $c(Q)$ that could be associated to the
$\mu$- process. The fit provides $\tau_\alpha(Q)$ to be $0.30 \pm
0.05$ ps at low $Q$, a value consistent with the hydrodynamic
limit $\tau(0)= 0.25\pm 0.05$ ps as calculated through the
relation \cite{PRE6055051999}:
\begin{equation}
\tau({0}) = {\nu_L - {\Gamma\mu (0)\over 2Q^2} \over c_{\infty
\alpha}^2(0) - c_0^2(0)}
\end{equation}

\noindent where $\nu_L$ and $c_0$ are respectively the kinematic
longitudinal viscosity and the adiabatic sound velocity measured
by Brillouin light scattering \cite{dapubblicare}. This numerical
equivalence gives further support to the validity of the employed
viscoelastic model.

Recent molecular dynamics simulation studies on HF
\cite{PRL8120801998, JCP11290252000, PRL8448782000} provide the necessary information to discuss
the present experimental results in the framework of the two
relaxation processes scenario. Similarly to Fig.~1 for liquid
water, in Fig.~5 we report the sound velocities of Fig.~4 together
with numerical simulation results \cite{PRL8120801998}: {\it i)}
$c_o(Q)$ as derived from the simulated static structure factor,
{\it ii)} $c(Q)$ as derived from the maxima of the simulated
longitudinal current spectra, and {\it iii)} $c_{\infty}(Q)$ as
derived from the fourth moment of the dynamic structure factor. It
is worth to note that the quantity $c_{\infty\alpha}(Q)$ does not
have a simple expression in terms of microscopic variables, and
cannot be directly evaluated numerically. In spite of the slightly
different thermodynamic points between the experiment (T=239 K)
and the simulation (T=203 K), we observe an excellent agreement
between the two common sets of data, i.~e. for $c_o(Q)$ and
$c(Q)$. This agreement implies that the interaction potential
model used in the simulation matches well the properties of the
real system or that these quantities are relatively insensitive to
the interaction potential. The important information emerging from
the comparison of the data in Fig.~5, is the very large difference
between the (measured) $ c_{\infty\alpha} = c_{o\mu}$ and the
(calculated) $c_{\infty}$. This, in turn, implies not only the
existence of the $\mu$-process, but also that, in HF, this process
has a relative strength substantially larger than in water over
the whole considered $Q$ range, which extends beyond the first
peak in the static structure factor.

In conclusion we provide a strong indication that, similarly to
simple and glass-forming liquids, also hydrogen bonded liquids
present {\it two} relaxation processes affecting their high
frequency collective dynamics. These two processes produce a
phenomenology consistent with the $\alpha$- and $\mu$- processes,
thus suggesting their universality in the liquid state. The
strengths of these two processes are, however, dependent on the
specific system. In particular - contrary to simple liquids where
$c_\infty/c_o \approx 1.2$ and $c_{\infty\alpha}$ differs from
$c_o$ by few percent \cite{JPC1280092000} - the strengths of these
relaxation processes are much larger in hydrogen bonded liquids:
at low $Q$ in both water and HF $c_{\infty\alpha} /c_o \approx$2,
while $c_{\infty}/ c_o \approx$3 in water and $\approx$7 in HF. It
is intriguing to understand the origin of these quantitative
differences, and to see whether they are correlated the different
hydrogen bond networks existing in water and HF.

\begin{acknowledgments}
We acknowledge D.~Fioretto for his help during the Brillouin light
scattering measurements, C.~Henriquet for the design, development and
assembly of the hydrogen-fluoride cell, C.~Lapras for technical
help and R.~Vallauri for useful discussions.
\end{acknowledgments}

%create the reference section using BibTeX:
%\bibliography{biblio}

\begin{thebibliography}{18}
\bibitem[{not()}]{nota1}
\bibinfo{note}{This ordering is based on the observation that $\tau_\alpha $ is
  always larger than $\tau_\mu$.}
\end{thebibliography}

\begin{figure}
\includegraphics[width=8.5cm,height=6.8cm]{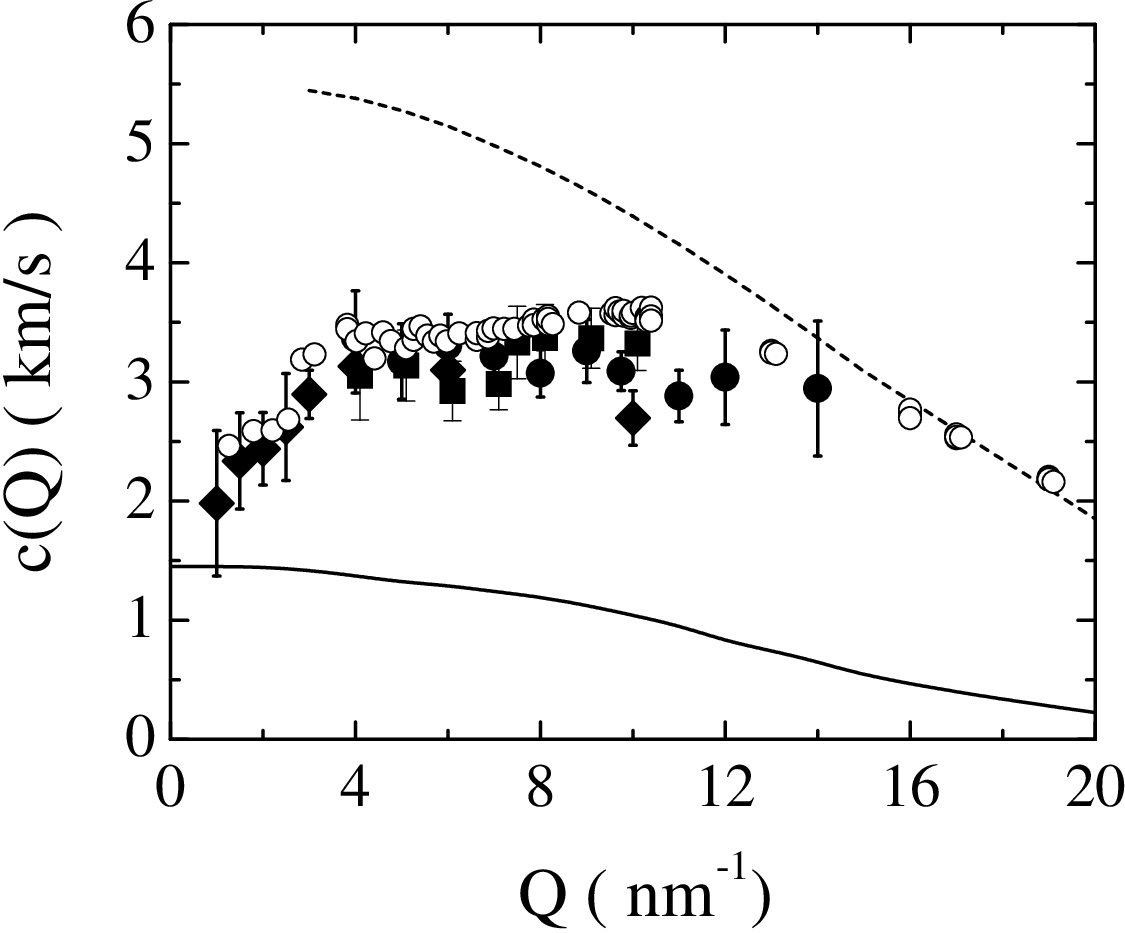}
\caption{Q dependence of experimental and theoretical velocities
of sound in water. Open circles, MD simulations
\cite{PRL7916781997}; full squares, circles and diamonds, IXS data
\cite{PRL77831996}; solid (dashed) line, zero (infinite) frequency
limit \cite{PRE4716771993}.} \label{fig1}
\end{figure}

\begin{figure}
\includegraphics[width=8.0cm,height=6.7cm]{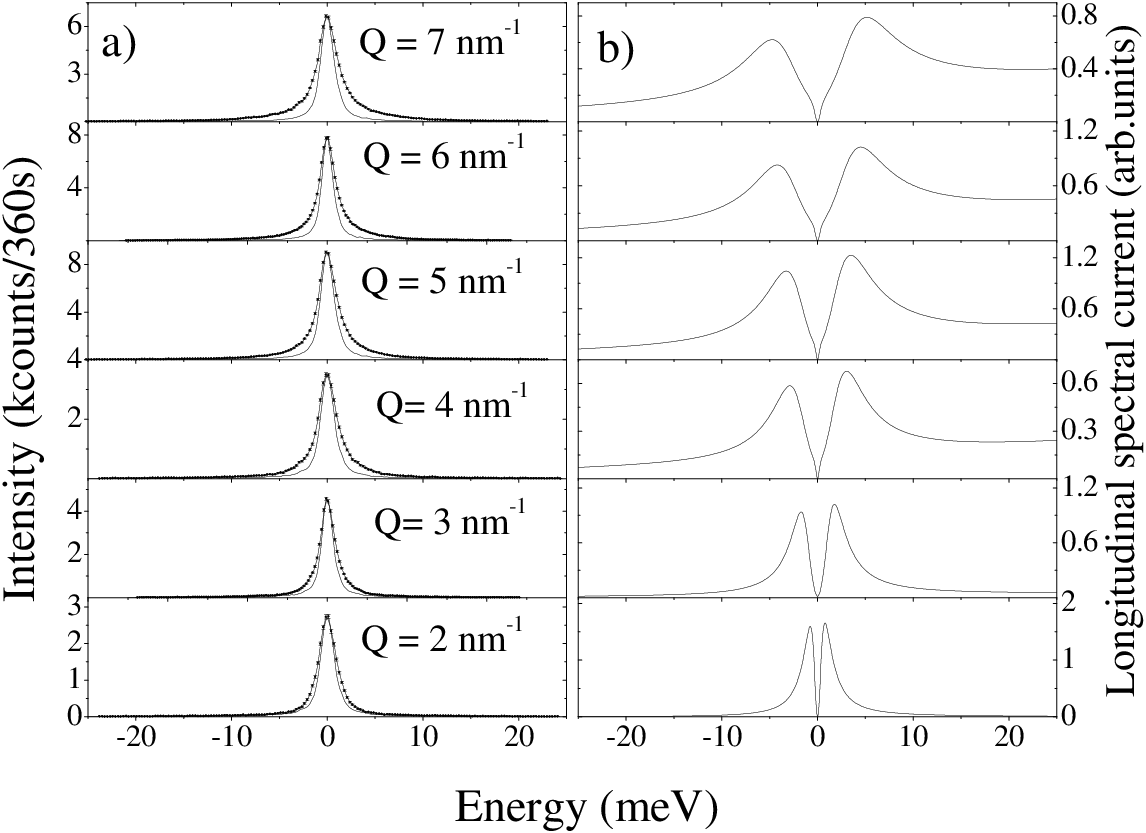}
\caption{ $a)$ IXS spectra of HF at fixed temperature $T = 239 K$
and at the indicated moment transfer Q plotted together with the
corresponding experimental resolutions (solid lines). $b)$ Resolution
deconvoluted inelastic part (DHO) of the longitudinal spectral
current.}
\label{fig2}
\end{figure}

\begin{figure}
\includegraphics[width=8.0cm,height=6.8cm]{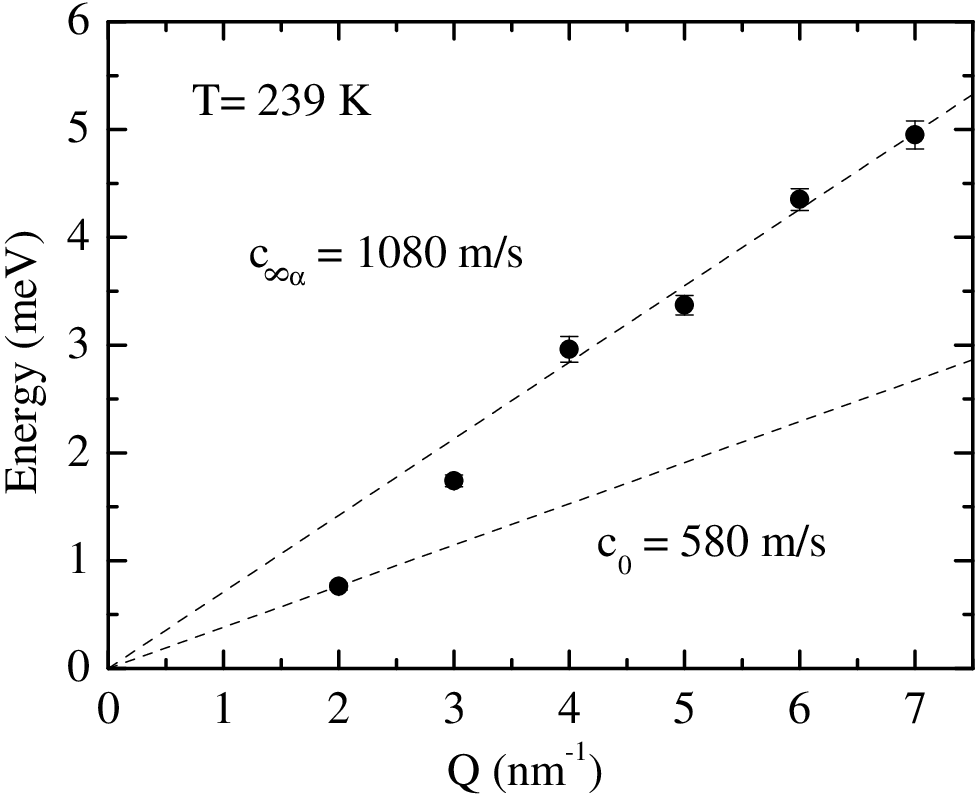}
\caption{Dispersion curve obtained from the maxima of the current
spectra shown in Fig.~\ref{fig2}. The upper dashed line is the linear fit to
the data, the lower dashed line indicates the adiabatic sound
velocity as measured by Brillouin light scattering \cite{dapubblicare}.
}\label{fig3}
\end{figure}

\begin{figure}
\includegraphics[width=8.3cm,height=6.9cm]{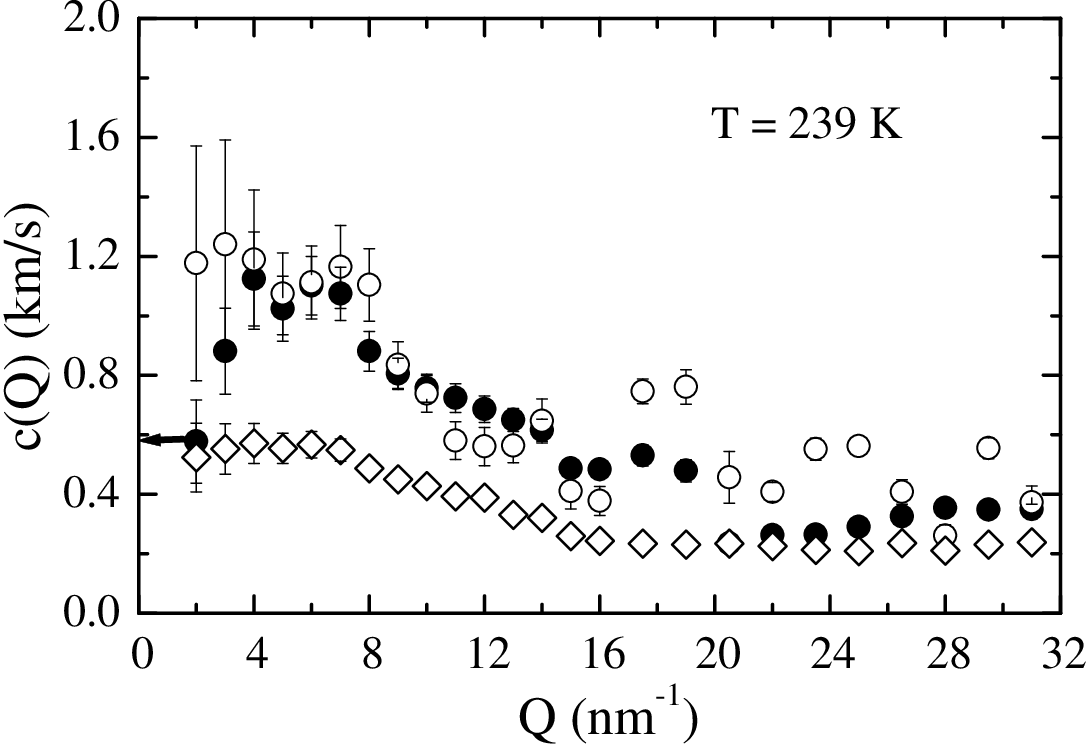}
\caption{Q-dependence of the sound velocities $c_0(Q)$ (open
diamonds) and $c_{ \infty \alpha}(Q)$ (open circles) from the
viscoelastic fit, and $c(Q)$ from Fig.~\ref{fig3} . The value of
the adiabatic sound velocity $c_0$ is indicated by the arrow
\cite{dapubblicare}.
}\label{fig4}
\end{figure}

\begin{figure}
\includegraphics[width=8.5cm,height=6.4cm]{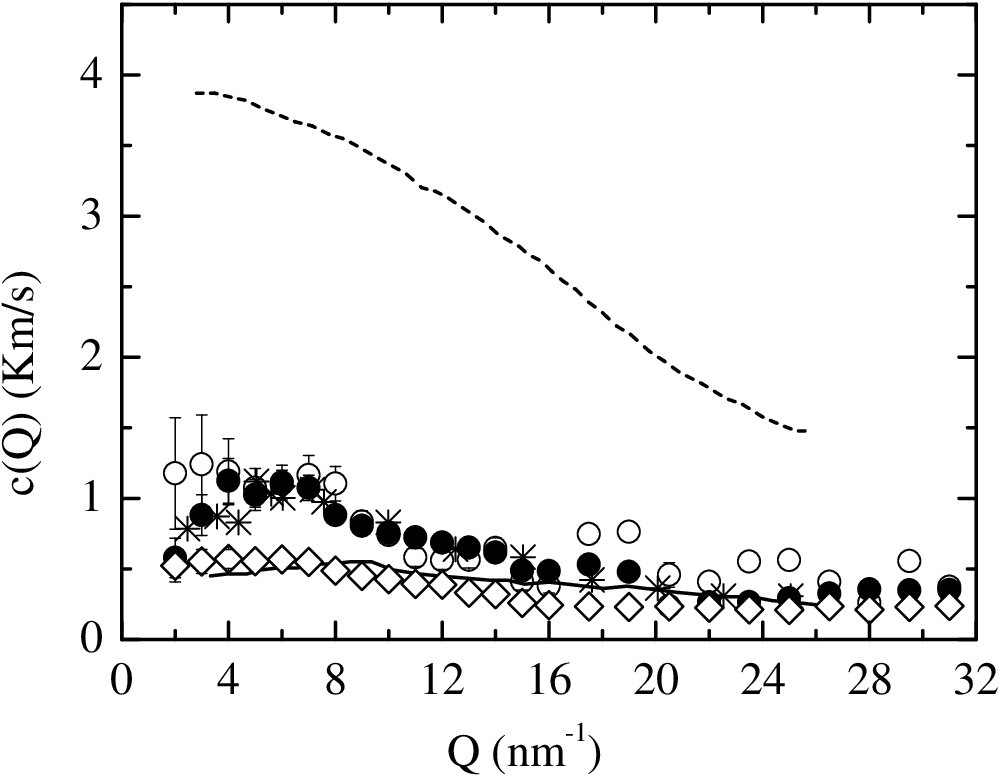}
\caption{Q-dependence of the sound velocities at $T$=239 K from
this work: $c_0(Q)$ (open diamonds), $c_{\infty \alpha}$ (open
circles), $c(Q)$ (full circles) and at $T$=203 K from the MD
results of Ref. \cite{PRL8120801998}: $c_0(Q)$ (solid line),
$c_{\infty \mu}$ (dashed line), $c(Q)$ (stars).
}\label{fig5}
\end{figure}

\end{document}